\def\preprint{0} 
\def\preprint{1} 
\def\comment#1{}
\if\preprint1
	\documentclass{article}
	\usepackage{mn2e}
	\usepackage{astrop-bib}
	\usepackage{times}
	\usepackage{graphicx}
\else
	\documentclass{article}
	\usepackage[referee]{mn2e}
	\usepackage{astrop-bib}
	\usepackage{times}
	\usepackage{graphicx}
\fi

\ifnfsstwo

\fi

\ifnfssone
  \newmathalphabet{\mathit}
    \addtoversion{normal}{\mathit}{cmr}{m}{it}
    \addtoversion{bold}{\mathit}{cmr}{bx}{it}

\fi

\ifoldfss

\fi

\loadboldmathitalic
\loadboldgreek

\title[Mode switching in R~Doradus]
{Mode switching in the nearby Mira-like variable R~Doradus}
\author[T.~R.~Bedding et al.]
       {T.~R.~Bedding,$^1$\thanks{E-mail: \tt bedding@physics.usyd.edu.au}
	Albert~A.~Zijlstra,$^2$ \thanks{E-mail: \tt aaz@aquila.phy.umist.ac.uk}
	A.~Jones$^3$ and 
	G. Foster$^4$\thanks{E-mail: \tt gfoster@aavso.org}\\
	$^1$School of Physics, University of Sydney 2006, Australia\\
	$^2$UMIST, Department of Physics, P.O. Box 88, Manchester M60 1QD,
		U.K.\\
	$^3$Carter Observatory, P.O. Box 2909, Wellington, New Zealand\\
        $^4$AAVSO, 25 Birch St., Cambridge, MA 02138, U.S.A.
}

\pubyear{1998}

\begin{document}

\maketitle

\begin{abstract}

We discuss visual observations spanning nearly 70 years of the nearby
semiregular variable R~Doradus.  Using wavelet analysis, we show that the
star switches back and forth between two pulsation modes having periods of
332 days and about 175 days, the latter with much smaller amplitude.
Comparison with model calculations suggests that the two modes are the
first and third radial overtone, with the physical diameter of the star
making fundamental mode pulsation unlikely.  The mode changes occur on a
timescale of about 1000\,d, which is too rapid be related to a change in
the overall thermal structure of the star and may instead be related to
weak chaos.

The Hipparcos distance to R~Dor is $62.4\pm2.8\,$pc which, taken with its
dominant 332-day period, places it exactly on the period-luminosity
relation of Miras in the Large Magellanic Cloud.  Our results imply first
overtone pulsation for all Miras which fall on the P-L relation.  We argue
that semiregular variables with long periods may largely be a subset of
Miras and should be included in studies of Mira behaviour.  The
semiregulars may contain the immediate evolutionary Mira progenitors, or
stars may alternate between periods of semiregular and Mira behaviour.
\end{abstract}


\section{Introduction}

Miras are large-amplitude, long-period variables located near the tip of
the Asymptotic Giant Branch (AGB).  Traditionally, stars are only
considered Miras if their peak-to-peak amplitude at $V$ exceeds 2.5\,mag.
The periods are typically between 200 and 500 days, although OH/IR stars (a
subset of the Miras which show large circumstellar extinction) have periods
up to 2000 days.  The periods are generally stable but the maximum and
minimum magnitude can vary from cycle to cycle.  Miras with periods longer
than 300 days often show evidence for high mass-loss rates.

Mira variability is associated with the thermal-pulsing AGB, where the star
alternates between periods of hydrogen and helium burning in a shell around
the inert carbon/oxygen core.  Mira pulsation occurs during the hydrogen
shell-burning phase, when the star is more luminous, although it is
possible that some stars also show Mira pulsations during the helium shell
flash (the `pulse'; \citebare{W+Z81}; \citebare{Zij95}).

\if\preprint1
\begin{figure*}

\centerline{ \includegraphics[draft=false,bb=86 341 541 703]
{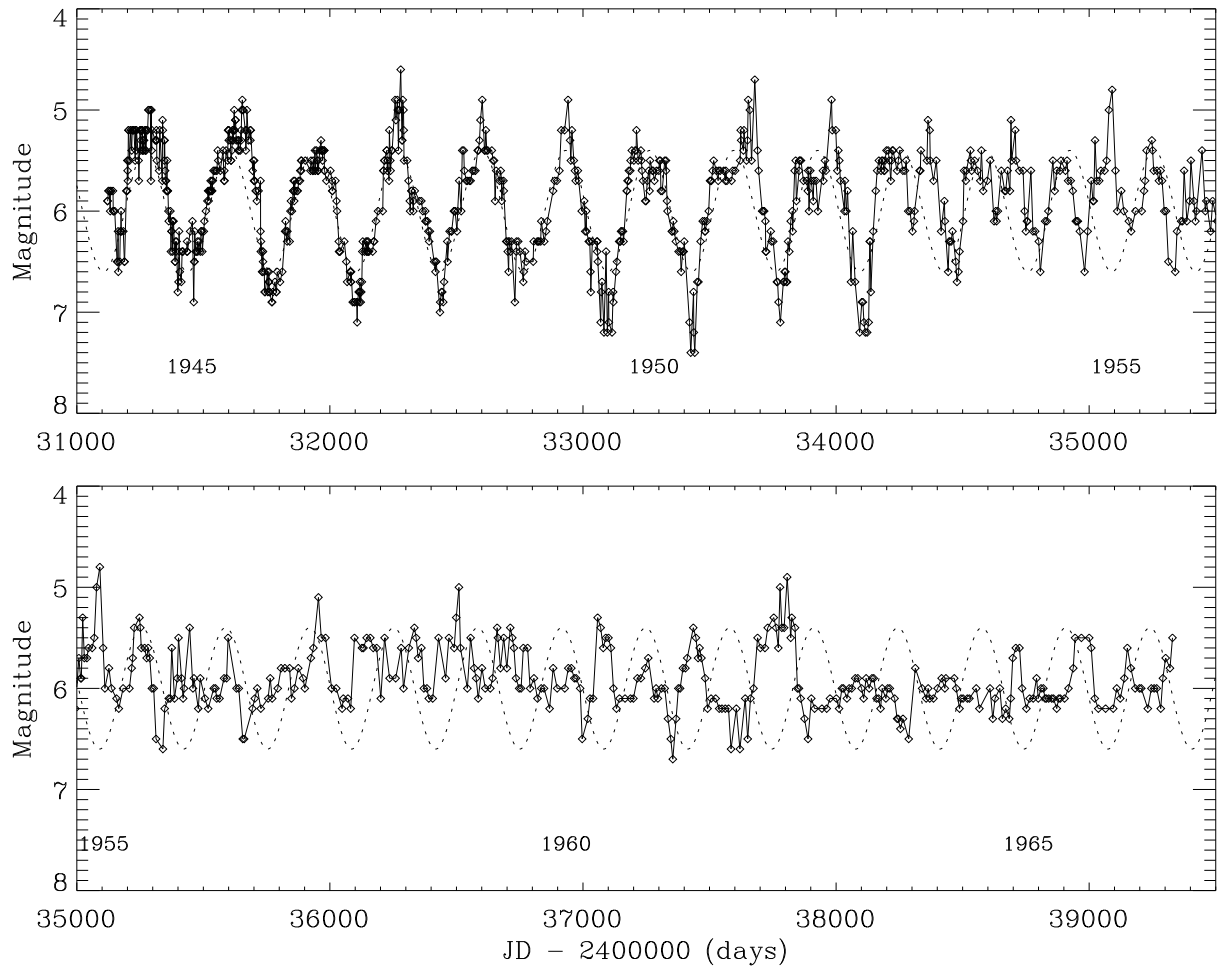}}

\caption[]{\label{fig.jones-light} Light curve of R~Dor (observations by
A. Jones).  The dashed curve shows a sinusoid with a period of 332 days.}

\end{figure*}

\fi

The existence of a well-defined and narrow period--luminosity (P-L)
relation for Miras in the LMC \cite{FGW89} is evidence that most Miras
pulsate in the same mode.  The identity of this mode, however, is still
controversial.  Temperatures and radii of Miras are consistent with first
overtone: direct measurements of stellar angular diameters indicate that
most Miras are larger and cooler than expected for fundamental mode
pulsators \cite{THB94b,Fea96,vLFW96}.  On the other hand, the observed
shock velocities in the CO lines are too large to easily be reconciled with
any mode other than the fundamental \cite{Woo90,HLS97}.

The semiregular variables differ from classical Miras in showing smaller
amplitudes ($< 2.5$ mag peak-to-peak) and/or less regular pulsations,
sometimes with multiple periods.  \citetwo{K+H92}{K+H94} have argued that
some stars classed as semiregulars are closely related to Miras, excluded
only because of the restrictive classical definition.  These stars could be
important for our understanding of Miras and could include their immediate
progenitors.

R~Doradus (HR~1492, M8\,III) is classified as a semiregular (SRb) with a
period of 338\,d \cite{Kho88}.  This period is within the normal range for
Miras, in contrast to most semiregulars, which have periods closer to 150
days.  We have recently measured the angular diameter of R~Dor to be
$57\pm5$ mas, the largest of any star except the Sun \cite{BZvdL97}.  We
argued that this star is closely related to the Miras, in spite of its more
complicated and smaller-amplitude variability, on the grounds that IRAS
images show extended dust emission centred on R~Dor \cite{YPK93b} and the
IRAS LRS spectrum shows a weak silicate feature \cite{V+C89}.  Both
indicate dust and mass loss, which are normally confined to Miras with
periods of more than 300 days.

Here we discuss visual observations of R~Dor spanning 70 years.  We find
that R~Dor switches between two different modes, one with a Mira-like
period and the other with a shorter period more typical of semiregulars.
The Hipparcos distance is used to show that the longer period fits the Mira
P-L relation.

\section{Observations by A. Jones}

One of us (AJ) has monitored R~Dor over a 23-year period (1944--1967),
producing about 1100 measurements (Fig.~\ref{fig.jones-light}).  R~Dor is
circumpolar from New Zealand, so there are no yearly gaps in the time
series.  Although fainter stars are observed using a home-made 317\,mm
$f/5$ Newtonian reflector, brighter stars such as R~Dor are observed using
a smaller finder telescope.  The magnitudes are estimated by visual
comparison with fields of bright standard stars.  When making visual
estimates, certain precautions were taken to eliminate possible errors
which are particularly important for a star as red as R~Dor.  Observations
of red stars are not made in conditions of bright moonlight.  When
observing a red star it is advisable to make quick glances, otherwise
visual observations may overestimate its brightness.  To make estimates,
each star is brought to the centre of the field before noting its
brightness.  If the variable and a comparison star are not far apart, the
observer's head is oriented so that the line joining the stars is parallel
to the eyes.  Another issue to be wary of is that the star nearest the
observer's nose may seem a little brighter than is the case.  When going to
the telescope, every effort is made not to recall any previous measurements
-- the observations reported in this paper have never been plotted by the
observer.

\if\preprint1
\if\preprint0
	\twocolumn
\fi
\begin{figure}

\centerline{ 
\includegraphics[draft=false, bb=63 345 278 719, width=\the\hsize]
{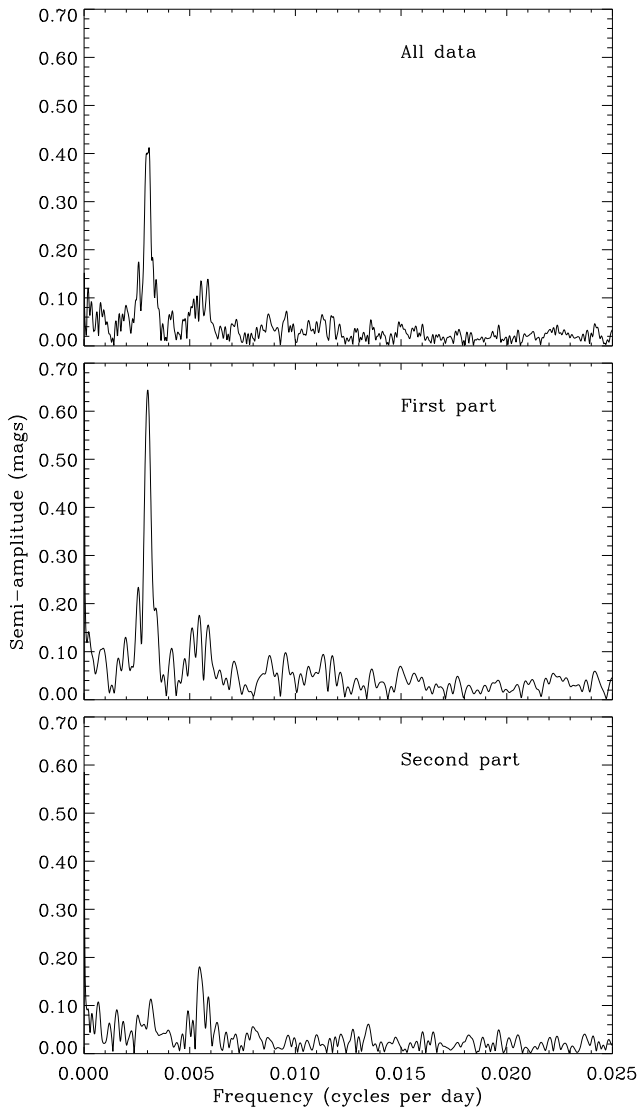}}

\caption[]{\label{fig.jones-amp} Amplitude spectra of the data in
Fig.~\ref{fig.jones-light}.  The data were taken as a whole (top panel),
and also in two parts (middle and lower panels), where the division was
made at JD\,2434500.   }

\end{figure}

\if\preprint0
	\onecolumn
\fi

\fi

Inspection of the data in Fig.~\ref{fig.jones-light} shows that the
pulsation behaviour of R~Dor changed significantly over time.  The dashed
curve shows a sinusoidal period of 332\,d.  The peak-to-peak amplitude of
the pulsation reached up to 1.5\,mag during the late 1940s, after which the
variation was much more irregular and of lower amplitude.  The presence of
variations with a shorter period can be seen, particularly around JD
2434000.

The Fourier amplitude spectra are shown in Fig.~\ref{fig.jones-amp}, with
the data taken both as a whole and also as two subsets.  There are two
strong periods of 332 days and about 175 days, with the longer period being
almost absent in the second part of the time series.

\if\preprint1
\begin{figure*}

\centerline{
\includegraphics[draft=false, bb=36 122 525 512]
{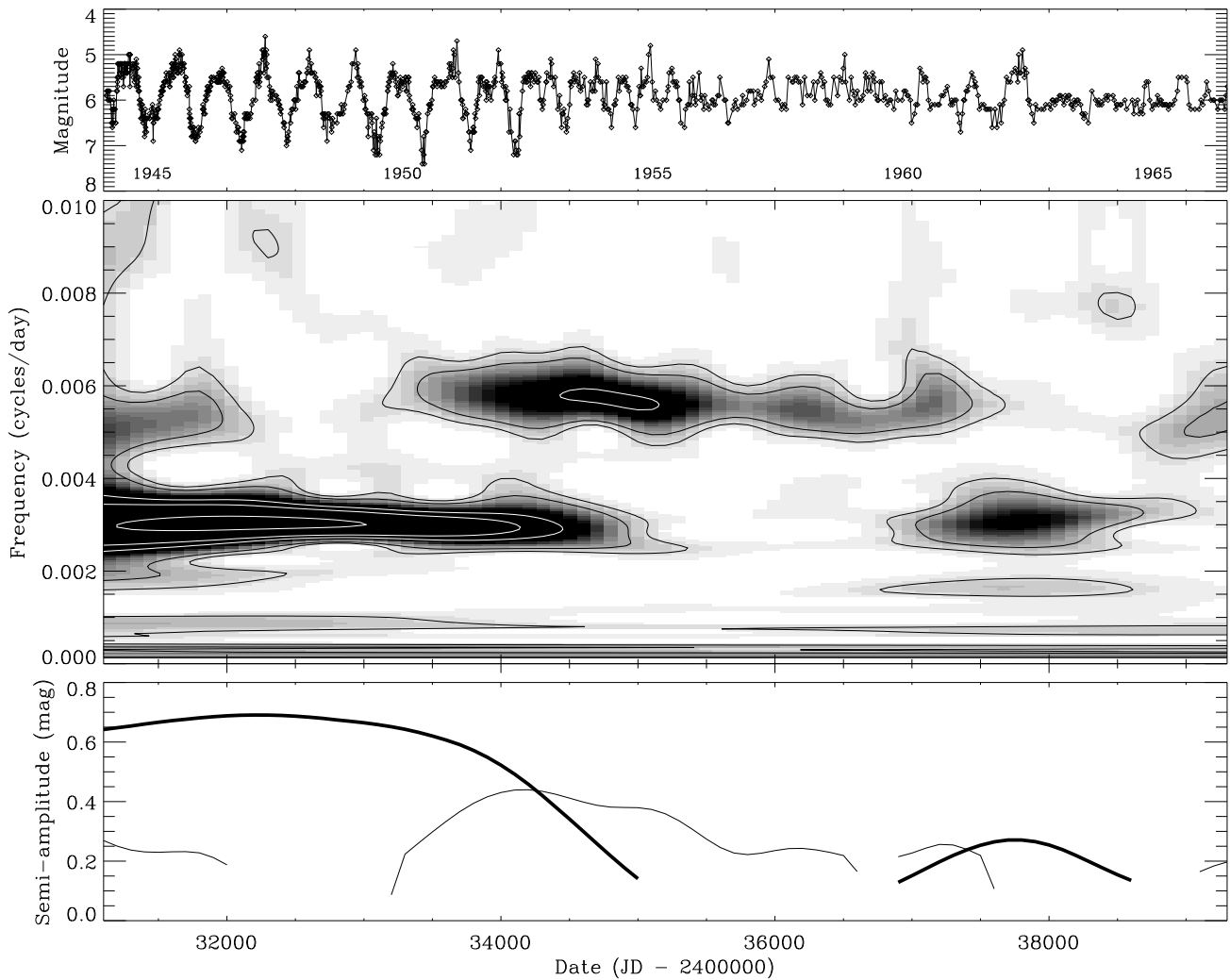}}

\caption[]{\label{fig.wwz.jones005} Wavelet analysis of observations of
R~Dor by A. Jones.  The top panel shows the light curve reproduced from
Fig.~\ref{fig.jones-light}.  The middle panel shows the WWZ and the bottom
panel shows the WWA evaluated at periods of 332\,d (thick line) and 175\,d
(thin line).  See text for more details. }

\end{figure*}

\fi

\subsection{Wavelet analysis}

We have also analysed the observations using wavelets.  This technique has
been shown by several groups to be a useful tool for investigating period
and amplitude changes in long period variables
\cite{S+V92,SVG94,G+S95a,SGK96,Fos96,B+M97}.  We have used the weighted
wavelet Z-transform (WWZ) developed by \citeone{Fos96} specifically for
unevenly sampled data.  We experimented with different values for the
parameter $c$, which defines the tradeoff between time resolution and
frequency resolution \cite{Fos96}, and settled on $c=0.005$ as a good
compromise.

The results are displayed in Fig.~\ref{fig.wwz.jones005}.  The top panel
shows the light curve, while the middle panel shows the WWZ (the bottom
panel is discussed below).  The two periods are clearly present, with
periods of 332\,d and 175\,d, and the star appears to alternate between
them.  However, care must be taken in interpreting the WWZ if we wish to
distinguish between mode switching, in which one period essentially
replaces the other, and the case in which one period decreases in amplitude
while the other remains roughly constant.  As discussed by \citeone{Fos96},
the WWZ is an excellent locator of the signal frequency but a poor measure
of amplitude.  He recommends estimating the amplitude using the weighted
wavelet amplitude (WWA), which is excellent for this purpose once the
signal frequency is known, but is less good at finding the correct
frequency.

\if\preprint1
\begin{figure*}

\centerline{
\includegraphics[draft=false, bb=36 122 525 512]
{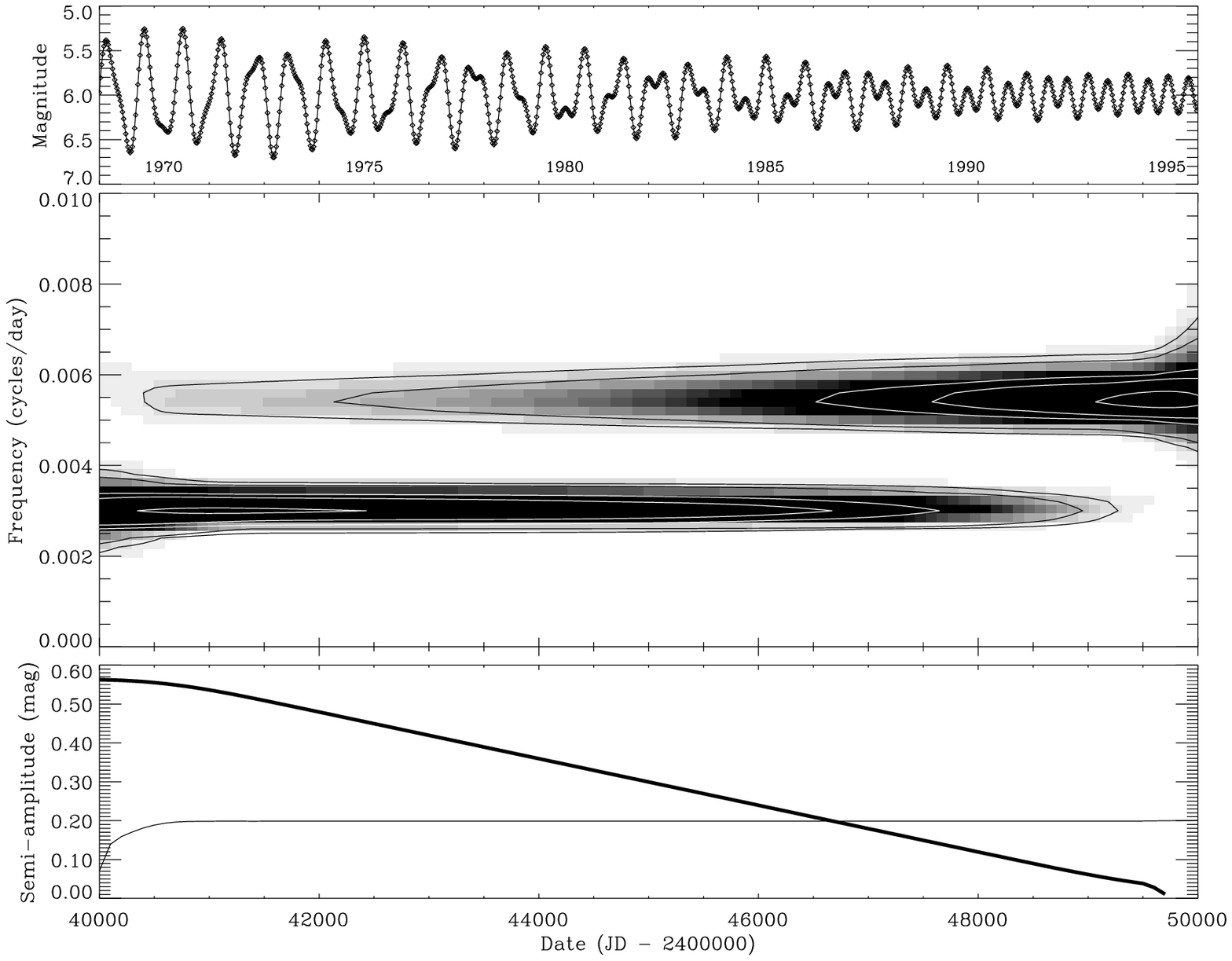}}

\caption[]{\label{fig.sim} Analysis of simulated data.  The light curve
(top) consists of a long-period component that decays linearly with time
and a short-period component with constant amplitude.  The WWZ (middle)
shows both periods, but give the false impression that the short-period
component is gradually increasing in strength.  Evaluating the WWA (bottom)
gives the correct amplitude behaviour at both the long period (thick line)
and the short period (thin line).}

\end{figure*}

\fi

To examine this, we have conducted a simulation which is shown in
Fig.~\ref{fig.sim}.  The light curve (top panel) is the sum of two
sinusoids with periods of 332\,d and 183\,d (the shorter period is slightly
different to that in R~Dor because the simulation was done before the real
period was confirmed).  The long-period signal has a semi-amplitude that
decreases linearly from 0.6\,mag to zero through the course of the time
series.  The shorter-period component has a constant semi-amplitude of
0.2\,mag.  The WWZ (middle panel) shows the two periods clearly.  At the
longer period the WWZ decreases with time, as would be expected for a
decaying signal, but at the shorter period the WWZ shows an {\em
increase\/} with time, despite the fact that the input signal has constant
amplitude.  The reason is that the WWZ essentially measures the {\em
significance\/} of the period in question, which naturally increases as the
other period becomes less dominant.  The bottom panel shows the
semi-amplitude at each of the two periods, calculated using the WWA\@.  We
see that, except for end effects, the WWA correctly recovers the amplitudes
of the input signals.  If we had relied on the WWZ alone, we might have
inferred a gradual mode switch, whereas the WWA correctly shows that there
has been no exchange of power.

Returning to R~Dor, the WWA at each of the two periods is shown in the
bottom panel of Fig.~\ref{fig.wwz.jones005}.  Unlike the simulation, there
is an exchange of power.  The first switch occurs around JD\,2434000, with
simultaneous strengthening of the shorter period and weakening of the
longer one.  There is a less prominent reverse switch starting at about
JD\,2437000 which is also apparent in the raw light curve, possibly
followed by another switch at the very end of the time series.

\if\preprint1
\begin{figure*}

\centerline{ 
\includegraphics[draft=false,bb=80 125 556 638]
{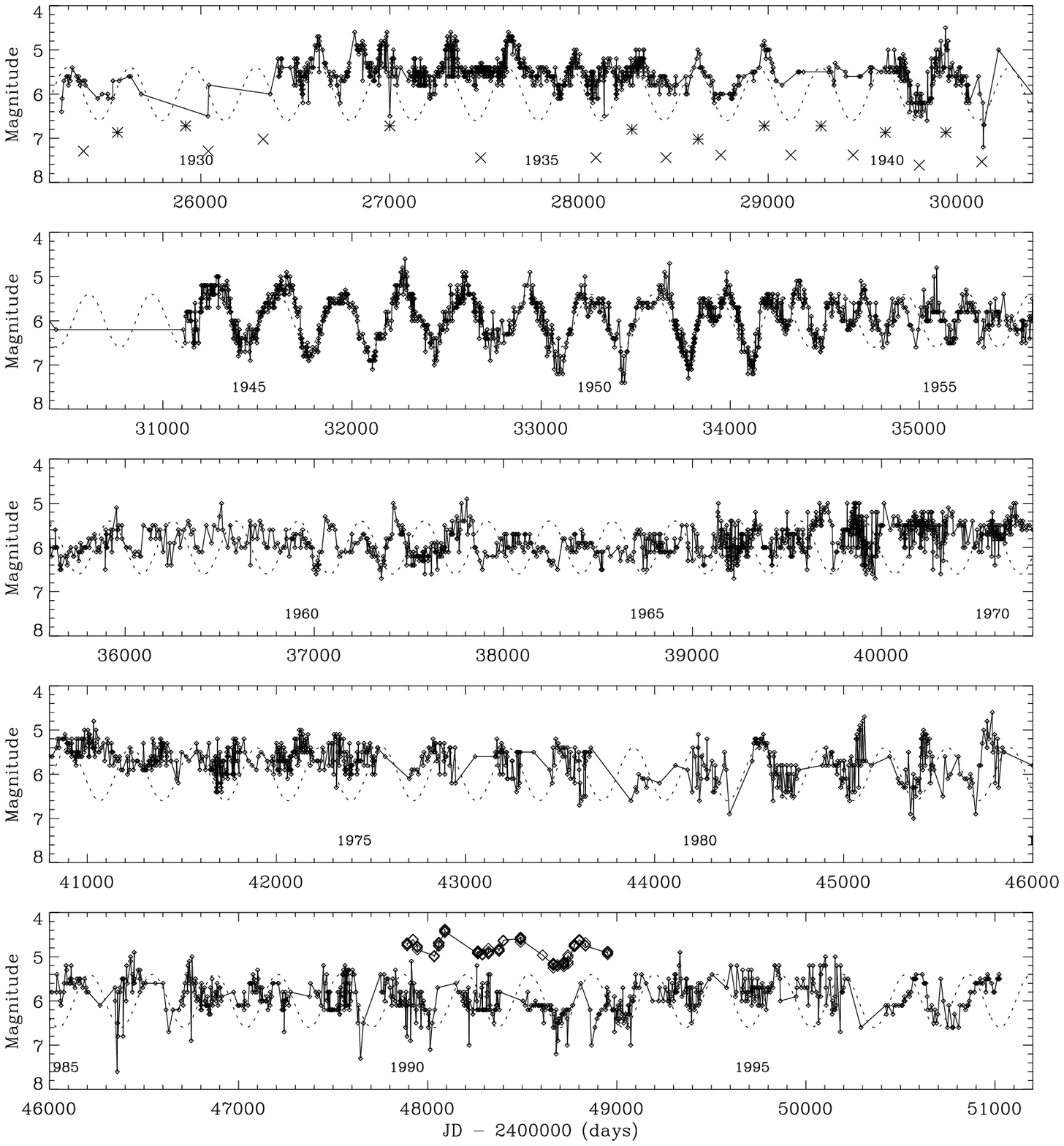}}

\caption[]{\label{fig.all-light} Light curve of R Dor from RASNZ data
(including observations in Fig.~\ref{fig.jones-light}).  In the bottom
panel, the diamonds are Tycho $VT$ data, shifted upwards by 1.0\,mag for
readability.  In the top panel, asterisks are maxima and crosses are minima
(in photographic magnitudes, shifted upwards by 0.5\,mag), reported by
\citeone{Gap52}.  Finally, note that all observations after JD\,2450400
were made by A.~Jones.  The dotted sine curve has the same parameters as
that in Fig.~\ref{fig.jones-light}.}

\end{figure*}

\fi

\section{Other observations}

Further investigation of R~Dor requires more data.  A. Jones has recently
resumed regular observations of R~Dor and these are presented below, but
there remains a gap of 30\,yr which can only be filled by other observers.
The archives of the Royal Astronomical Society of New Zealand (RASNZ)
contain about 4000 measurements from other observers, which we have
analysed.  The largest individual contributions are 660, 400, 250, 340,
190, 170, 126, 124 and 123 measurements.  There are 12 observers who
contributed 50--100 measurements and 19 who contributed 20--50.  Data from
one of these observers were discarded because they were systematically
fainter by 1\,mag than the local trend, illustrating the variations between
observers for extremely red stars and highlighting the desirability of
relying on a single observer.  Data from observers who contributed fewer
than 20 measurements were not included.

The combined light curve, including the measurements by AJ (both old and
new), is shown in Fig.~\ref{fig.all-light}.  Also shown in the top panel
are photographic magnitudes at maximum and minimum reported by
\citeone{Gap52}, upon which the catalogued period is presumably based and
which match the visual data quite well.  We also show 96 photometric
measurements by the Tycho instrument on the Hipparcos mission
\cite{ESA97,HBB97}.  The accuracy of these space-based observations is much
higher than the visual data, but they only span four years.

\if\preprint1
\begin{figure*}

\centerline{
\includegraphics[draft=false,bb=36 122 525 512]
{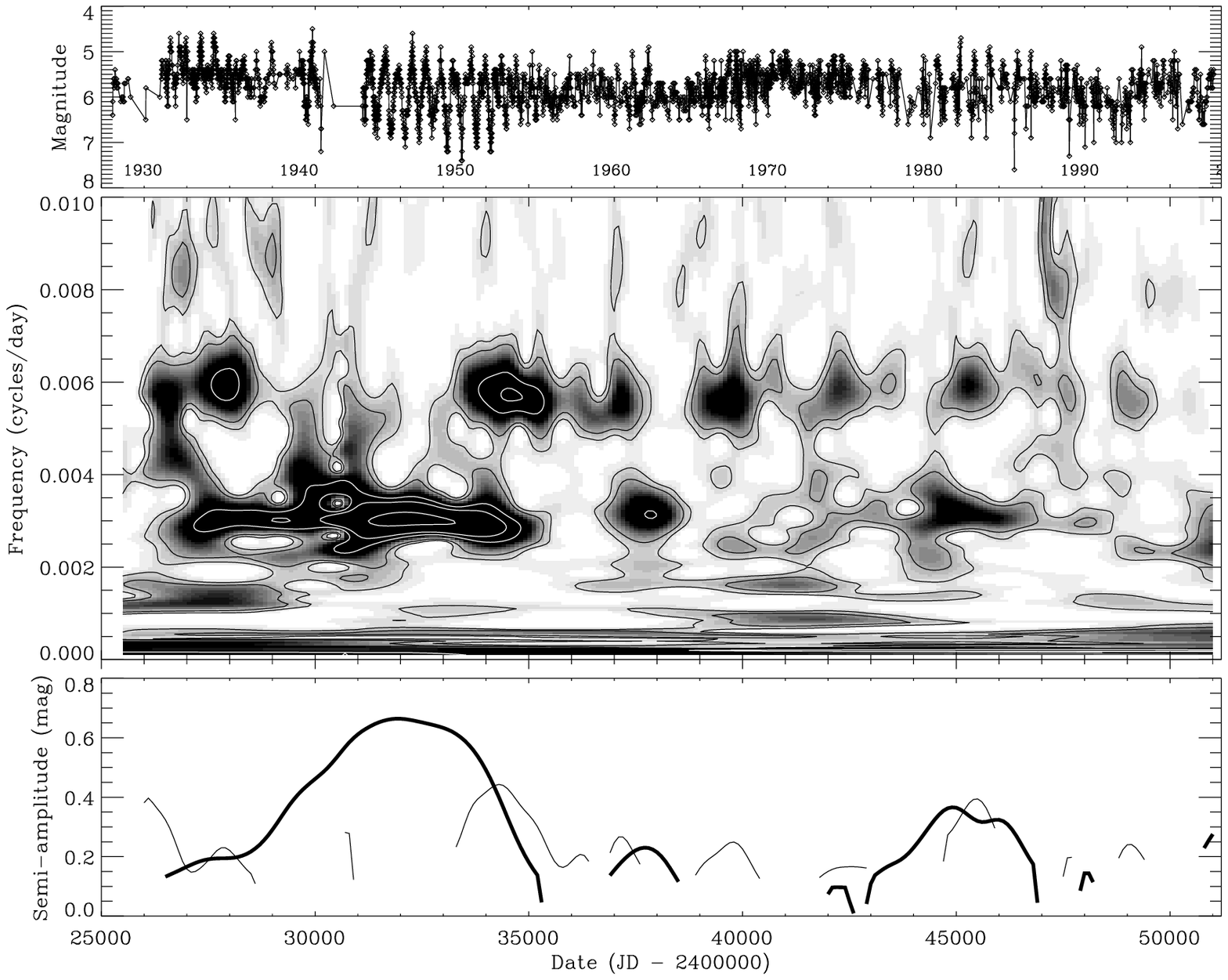}}

\caption[]{\label{fig.wwz.all005} Wavelet transform for R~Dor.  Same as
Fig.~\ref{fig.wwz.jones005}, but based on the data in
Fig~\ref{fig.all-light}.}

\end{figure*}

\fi

The wavelet analysis of the full data set (including the Tycho data but
excluding the photographic extrema) is shown in Fig.~\ref{fig.wwz.all005}.
Both periods recur throughout the series and more mode switches are seen.
The change between the different types of pulsation occurs on a time scale
of about 1000 days.

\section{Relation to the Mira P-L relation}

On the basis R~Dor is closely related to the Mira variables, we estimated
the distance in \citeone{BZvdL97} by fitting the catalogued period (338\,d)
to the LMC Mira P-L relation of \citeone{Fea96}.  Our result of
$61\pm7\,$pc is now seen to be in excellent agreement with the Hipparcos
distance of $62.4\pm2.8\,$pc (parallax = $16.02 \pm 0.69\,$mas;
\citebare{ESA97}).  No true Miras lie within about 100\,pc and none has a
distance measurement as accurate as that of R~Dor.

Table~\ref{tab.RDor.Data} lists fundamental data for R~Dor, derived using
the Hipparcos distance and observations listed in \citeone{BZvdL97}.  The
uncertainty in the stellar radius is dominated by the error in the measured
angular diameter.  The uncertainty in the luminosity comes equally from the
uncertainties in $m_{\rm bol}$ and the distance.

\if\preprint1
	\begin{table}
\caption[]{\label{tab.RDor.Data} Fundamental parameters of R~Dor}

\begin{tabular}{ll}
\hline
\noalign{\smallskip}
Distance	&  $62.4\pm2.8$\,pc	\\
Radius   	& $383\pm38$\,R$_\odot$	\\
Effective temperature 	& $2740\pm190$\,K	\\
Luminosity 	& $6830\pm900$\,L$_\odot$	\\
$M_{\rm bol}$ 	& $-4.87 \pm 0.13 $		\\
Periods 	& 332 and $\sim$175 days \\
\noalign{\smallskip}
\hline
\end{tabular}
\end{table}

\fi

\if\preprint1
\begin{figure}

\centerline{ 
\includegraphics[width=7cm,bb=32 37 288 741,clip]{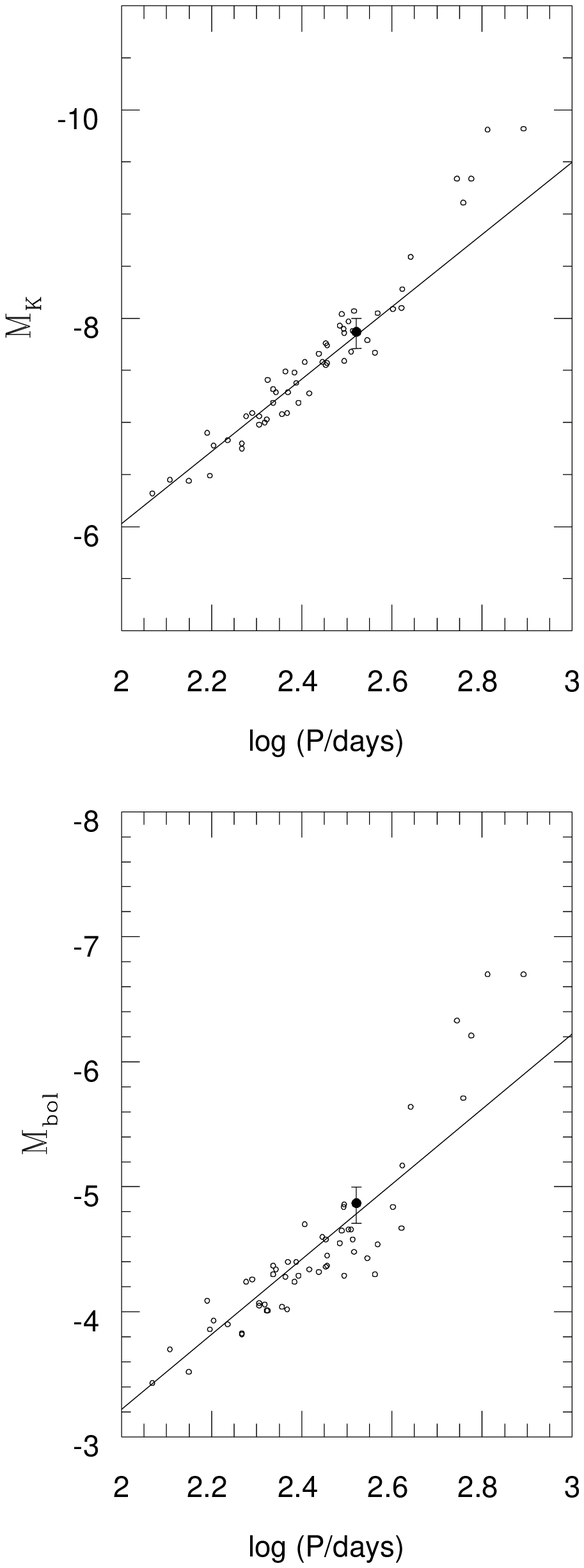}}

\caption[]{\label{fig.mirapl} The position of R~Dor (filled circle)
relative to the P-L relation for LMC Miras.  }

\end{figure}

\fi

Using these data, we can plot the position of R~Dor relative to the Mira P-L
relation.  Figure~\ref{fig.mirapl} shows the result, where the lines
indicates the P-L relations as given by \citeone{Fea96}:
\begin{eqnarray}
  M_K~ & = & -3.47 \log P + 0.91 \\
  M_{\rm bol} & = & -3.00 \log P + 2.78,
\end{eqnarray}
where $P$ is the period in days.  The small circles give the LMC data on
which the relation is based \cite{FGW89}, assuming a distance modulus to
the LMC of 18.56.  Note that the relation is better defined using the
$K$-band magnitude than the bolometric magnitude, probably due to
uncertainties in the bolometric corrections for such red stars.  The LMC
stars with periods longer than about 400 days are known to be over-luminous
with respect to the relation (see also \citebare{ZLW96}).

R~Dor is indicated by the filled circle, where the dominant 332-day period
was used.  The very small error bars leave no doubt that R~Dor fits the
relation extremely well.  We confirm the suggestion by \citeone{BZvdL97}
that, in spite of its semiregular behaviour, R~Dor can be considered a
Mira-like variable and its evolutionary state must be very closely related
to the regular large-amplitude Miras.

\section{Models for R~Doradus}

The internal structure of AGB stars, where most of the mass is centrally
concentrated and the radius is determined by a highly extended envelope,
leads to a large ratio between the periods of the fundamental pulsation
mode and the first overtone \citeeg{Woo75}.  It also leads one to expect
that a fundamental mode pulsator will have a much smaller radius than a
star with the same mass and luminosity that is pulsating in an overtone.
This difference has been used to distinguish between fundamental and
overtone pulsators for those (relatively few) Galactic Miras that have a
measured angular diameter.  By estimating distances using the LMC P-L
relation, \citeone{Fea96} concluded that the fundamental mode could be
excluded.  Using Hipparcos distances, \citeone{vLFW96} confirmed this
conclusion for most stars, and they found two whose radii appeared
consistent with fundamental mode (but see also \citebare{Bar98}).

In \citeone{BZvdL97} we performed this test for R~Dor using our measured
angular diameter and the distance from the P-L relation to estimate a mass.
We used the standard pulsation equation,
\begin{equation}
  Q = P \left({M}/{M_\odot}\right)^\frac{1}{2}
  \left({R}/{R_\odot}\right)^{-\frac{3}{2}},
\end{equation}
where $P$ is the period in days, $R$ and $M$ are the stellar radius and
mass in solar units and $Q$ is the pulsation constant (in days), which is
about 0.04\,d for the first overtone and 0.09\,d for the fundamental mode
\citeeg{F+W82}.

Using the revised radius and period in Table~\ref{tab.RDor.Data}, we can
repeat this calculation.  In first overtone, the stellar mass is found to
be $M= 0.8 \pm 0.25\,M_\odot$, while in fundamental mode we find $M= 4.1
\pm 1.4\,M_\odot$.  As in \citeone{BZvdL97}, we find that fundamental mode
pulsation implies a mass for R~Dor that is higher than expected from its
period \cite{Fea89,V+W93}.  For this reason, overtone pulsation is strongly
favoured.

The fact that R~Dor shows two well-defined periods may help in the mode
identification.  Table~\ref{tab.RDor.Wood} lists models calculated by P.
Wood (private communication) for the stellar parameters of R~Dor.  The
calculations assumed a core mass of 0.62\,$M_\odot$ and abundances of
helium and heavier element of $Y=0.30$ and $Z=0.02$, respectively.  The
first three rows show models that were calculated assuming that the primary
period (332\,d) corresponds to the fundamental mode.  For each model, a
mass was chosen and the ratio $l/H_p$ of the mixing length to the pressure
scale height was adjusted to give the correct period.  The resulting
effective temperature, together with the periods of the other modes ($P_i$)
and the mode growth rates per period ($GR_i$) and $Q$ values, are shown in
the table.  

The next three models were calculated in the same way, but assuming that
the primary period corresponds to the first overtone.  The growth rate for
the fundamental mode ($GR_0$) is very high in these models, implying that
the star would be unstable.  This can be taken as evidence against first
overtone pulsation, within the constraints of the model.

These models confirm that fundamental mode pulsation cannot occur at the
measured stellar radius unless the mass is greater than about 3\,$M_\odot$.
If we do identify the dominant period in R~Dor with the fundamental mode
then we must identify the observed secondary period with the first.  If we
instead identify the dominant period with overtone pulsation, the models
imply that the secondary period is third overtone or higher.

\if\preprint1
\newcommand{\m}{\llap{$-$}}

\begin{table*}
\caption[]{\label{tab.RDor.Wood} Theoretical models of Mira variables, showing
periods (in days) and growth rates per period}
\begin{flushleft}
\begin{tabular}{lcccrclrclrclrcl}
\hline
\noalign{\smallskip}
   ~$M$     & $T_{\rm eff}$  & $R$ & $l/H_p$   & 
  \multicolumn{3}{c}{Fundamental}  &
  \multicolumn{3}{c}{1st overtone} &
  \multicolumn{3}{c}{2nd overtone} &
  \multicolumn{3}{c}{3rd overtone} \\ 
\llap{(}$M_\odot$\rlap{)} & (K)   & \llap{(}$R_\odot$\rlap{)} &         &
  ~$P_0$~ & ~$Q_0$~ & $GR_0$~~ &
  ~$P_1$~ & ~$Q_1$~ & $GR_1$~~ &
  ~$P_2$~ & ~$Q_2$~ & $GR_2$~~ &
  ~$P_3$~ & ~$Q_3$~ & $GR_3$ \\
\noalign{\smallskip}
\hline
\noalign{\smallskip}
\multicolumn{16}{c}{Models with $L=6800\,L_\odot$ (P. Wood, private communication)}\\
\noalign{\smallskip}
  1.5 & 3138  & 279 & 2.68 & 332.7  & 0.087 &   0.11  & 154.7 & 0.041 & 0.20   &  120.2 & 0.032 & \m0.009  &  96.9 & 0.025 &   0.054  \\
  2.0 & 2970  & 312 & 2.00 & 334.6  & 0.086 & \m0.030 & 159.1 & 0.041 & 0.12   &  107.8 & 0.028 &   0.028  &  96.5 & 0.025 &   0.030  \\
  3.0 & 2787  & 354 & 1.39 & 334.0  & 0.087 & \m0.036 & 155.8 & 0.040 & 0.052  &   97.5 & 0.025 &   0.086  &  83.5 & 0.022 & \m0.021  \\
\noalign{\medskip}				      			       			       
  0.8 & 3038  & 298 & 3.41 & 601.4  & 0.105 &   3.5   & 332.1 & 0.058 & 0.0051 &  227.3 & 0.039 &   0.20   & 190.8 & 0.033 &    0.44  \\
  1.0 & 2776  & 357 & 2.54 & 779.2  & 0.115 &   2.7   & 332.5 & 0.049 & 0.0012 &  259.3 & 0.038 &   0.30   & 219.6 & 0.033 &    0.10  \\
  1.5 & 2387  & 483 & 1.48 & 1276.3 & 0.147 &   1.24  & 335.5 & 0.039 & 0.30   &  276.0 & 0.032 & \m0.0076 & 234.0 & 0.027 &  \m0.013 \\
\noalign{\smallskip}							       
\hline									       
\noalign{\smallskip}							       
\multicolumn{16}{c}{Model with $L=5775\,L_\odot$ \cite{B+T94}}\\
\noalign{\smallskip}							       
  1.15 & 2295 & 482 & 1.2  &
 1658.3 & 0.169 & 1.68\rlap{$^a$} &  
  334.6 & 0.034 & 0.50\rlap{$^a$} &  
  236.7 & 0.024 & 0.036\rlap{$^a$}&  
  176.1 & 0.018 & 0.19\rlap{$^a$} \\ 
\noalign{\smallskip}
\hline
\end{tabular}
{\footnotesize
\\
$^a$growth rates in \citeone{B+T94} are per day and have been converted to
per cycle.\\ 
}

\end{flushleft}

\end{table*}





\fi

The final row in Table~\ref{tab.RDor.Wood} is a model calculated by
\citeone{B+T94} for the Mira variable S~CMi.  Coincidently, the period of
this star (334\,d) is very close to that of R~Dor.  The model was
calculated assuming that the dominant period corresponds to the first
overtone.  The model luminosity, while somewhat lower than that of R~Dor,
lies only just outside our estimated ($1\sigma$) uncertainty range.
Interestingly, the period of the third overtone in the model is in
excellent agreement with our observed secondary period in R~Dor.  Better
agreement is possible because the period ratios (i.e., the $Q$ values) are
significantly different from those of Wood.  The model growth rates also
imply that the first and third overtones should be stronger than the
second.  Note, however, that the effective temperature of this model is
significantly lower than that of R~Dor (and hence the radius is larger), so
it is certainly not an ideal model.  Finally, we note that models by
\citeone{B+M97} for $o$~Cet give similar period ratios.  The difference
between these models and those of Wood resides in the choice of mixing
length, which is a matter of some uncertainty.

In summary, while we cannot definitely identify the pulsation modes in
R~Dor, the model of \citeone{B+T94} reproduces the two measured periods
very well as the first and third overtones.  Since the dominant period of
R~Dor places it exactly on the Mira P-L relation, our results imply that
this relation most likely corresponds to first overtone pulsation.

\section{Discussion}

\subsection{Other cases of mode switching}

Mode switching in a true Mira variable has not been observed.  There have
been reports of one or more secondary periods in the following Miras: S~CMi
and $\chi$~Cyg \cite{B+T94}, BS~Lyr \cite{Man96} and $o$~Cet \cite{B+M97}.
However, the secondary periods have much smaller amplitudes than the
dominant period, making it difficult to be certain of their reality.
\citeone{MFH97} recently reported secondary periods in two Miras (T~Col and
T~Eri) and in both cases the period ratio is 1.9.

Among semiregulars, several definitely show two periods.  Note that the
star Z~Aur, whose period changes were reported by \citeone{Lac73} and
discussed by \citeone{Woo75}, is actually a yellow supergiant of the
RV~Tauri class of variables \cite{Joy52} and is not relevant to this
discussion.  \citeone{CWS91} found evidence for mode switching in three
semiregular variables (RV~And, S~Aql and U~Boo).  They suggested that these
stars appear to switch between large-amplitude, long-period Mira-like
oscillations and short-period, lower-amplitude ones, as we have found for
R~Dor.  \citeone{P+D96} observed mode switching in the SRb star W~Boo from
25\,d to 50\,d\@.  \citeone{SGK96} list ten semiregulars that have two
periods and \citeone{G+S95} suggest that some of these have visual light
curve variations that may be connected with mode switching.  Most recently,
\citeone{MFH97} listed 28 semiregulars with two periods (4 in common with
the list of \citebare{SGK96}) and suggested that multiple periods are
common in semiregulars.  Period ratios all lie within the range 1.7 to 2.0
and R~Doradus, with a ratio of 1.81, is no exception.  We note that this
narrow range of ratios is consistent with the conclusion that all these
stars are pulsating in the same pair of modes.

The star most similar in behaviour to R~Dor is probably V~Boo.
\citeone{SGK96} discuss the visual light curve of this star in some detail
and show that the amplitude of the primary mode (period 258\,d) has
decreased with time.  They discuss the secondary period (137\,d), which
they say has constant amplitude, leading them to conclude that mode
switching has not occurred.  However, inspection of the light curve (their
Fig.~1) reveals similarities to R~Dor: a strong long period which then
gives way to double peaks of smaller amplitude.

\if\preprint1
\begin{figure*}

\centerline{
\includegraphics[draft=false,bb=36 122 525 512]
{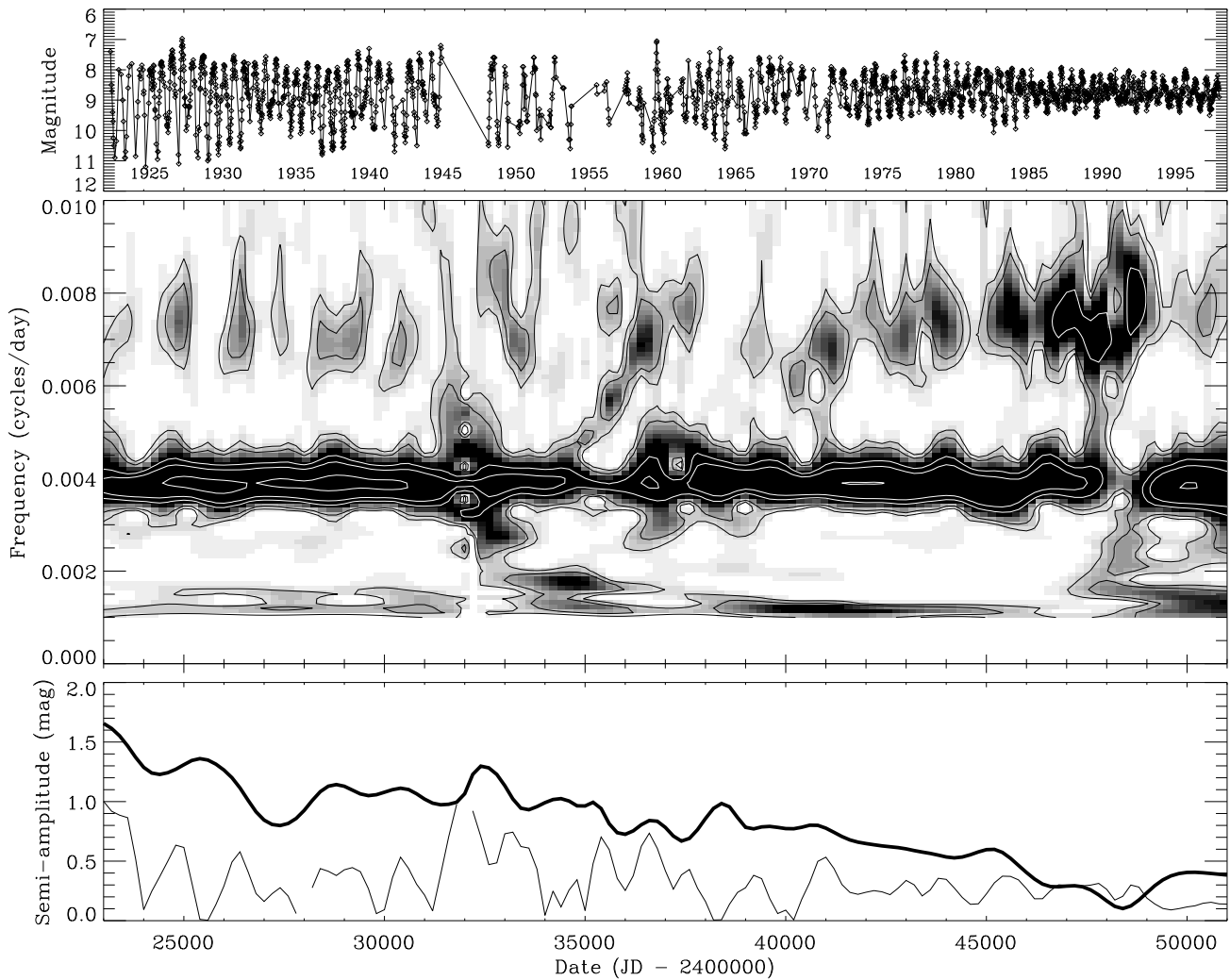}}

\caption[]{\label{fig.vboo} Wavelet transform for V~Boo based on the AFOEV
and VSOLJ data.  Same layout as Fig.~\ref{fig.wwz.jones005}, except that
the periods in the lower panel (WWA) are 258\,d (thick line) and 137\,d
(thin line).}

\end{figure*}

\fi

We have analysed visual measurements of V~Boo made by the AFOEV
(Association Francaise des Observateurs d'Etoiles Variables) and the VSOLJ
(Variable Stars Observers League in Japan).  This is a subset of the data
analysed by \citeone{SGK96}, except that the database now extends to
JD~2451000, giving an extra 1000 days of measurements.  Data from observers
who contributed fewer than 50 measurements were not included.  The
observations (9456 in total) were binned into 5-day averages, yielding 2551
data points.  The light curve and wavelet analysis are shown in
Fig.~\ref{fig.vboo}.  The gradual decrease in amplitude of the primary
period is confirmed.  The amplitude of the secondary period, while
variable, does not correlate (or anti-correlate) very strongly with that of
the primary period through most of the time series, although there is an
indication of mode switching near the end.  It appears that V~Boo has
changed from a true Mira to a semiregular, as suggested by \citeone{SGK96},
and may be entering a mode-switching phase similar to that currently being
shown by R~Dor.

Although there are still only a few reported cases of mode switching
behaviour, it may not be unusual.  Identification of mode switching in
variables with such long periods requires a long baseline of observations.
The fact that mode switching is seen in R~Dor, which is the nearest
Mira-like star, is consistent with it being fairly common.  Evolutionary,
periodic mode switching may occur immediately prior to the phase of fully
regular Mira pulsations.  Although the evolution from semiregular to Mira
has not been studied previously, the case of R~Dor suggests that this
change takes place over an extended time, and that mass loss begins well
before the change is completed.  Alternatively, R~Dor may belong to a
separate class of Mira-like variables which shows intermittent pulsation
behaviour.

\subsection{Models for mode switching}

Three theoretical models are available which include Mira mode switching.
\citeone{Kee70} has presented a model in which a Mira switches between the
fundamental and the first overtone and back again.  The change is gradual,
extending over tens of (overtone) periods, and is linked to a change in
luminosity.  

\citeone{Y+T96} discuss semi-stable models, where a seemingly stable
solution after several tens of cycles adjusts its internal energy structure
and therefore its pulsation behaviour.  In their non-linear models, the
star changes over a period of several hundred years (approximately the
thermal time scale of the envelope) from first overtone to a new
fundamental mode, where the latter has a shorter period than predicted from
the earlier overtone pulsation.  This result questions the usefulness of
mode analysis.  The model star pulsates for the first 300 years in an
overtone, after which the fundamental mode grows and becomes dominant.
During this time the fundamental mode changes in period from 508 to 330
days.

Both these models predict the change from overtone to fundamental to be
gradual, over tens of cycles, which is unlike the behaviour we see in
R~Dor.  It does not appear that the mode switching in R~Dor occurs on the
thermal time scale of the envelope, and the thermal structure of the star
can be assumed constant over the mode change.  We note that in both the
above models, the switch is between the first overtone and fundamental
mode, whereas in R~Dor it appears to be between first and third or higher
overtones.  The fact that R~Dor fits the Mira P-L relation so well strongly
argues against effects of non-linear growth.  Both the low pulsation
amplitude and the short mode life time in R~Dor may keep the star within
the linear regime.  This may argue against the \citename{Y+T96} model for
all Miras which agree with the P-L relation.

The third applicable model involves the effects of chaos on the stellar
structure and was developed by \citeone{IFH92}, based on work by, e.g.,
\citeone{B+G88}.  The model is described as a driven oscillator, with the
driving mechanism located well below the photosphere.  They show that the
pulsations becomes less regular as the envelope mass decreases, moving from
stable to multi-periodic to chaotic solutions.

The chaotic solution of \citeone{IFH92} corresponds to weak chaos, where
orbits associated with different modes intersect because of small
perturbations to the star.  The strongest effects are found near maximum
radius when the velocity in the atmosphere is near zero.  The star can
extend its stay here causing characteristic humps on the light curve near
maxima (\citename{IFH92}, their Figure 9).  This model appears to fit R~Dor
quite well.  In particular, the model depicted in the bottom panel of their
Fig.~13 shares features with R~Dor: the growing bump on the declining part
of the light curve, and the shift between modes on a short time scale.
Chaotic models can show very sudden changes in oscillation characteristics,
sometimes after a long sequence of regular oscillations.  The presence of
chaotic variations may therefore be common to Miras.

It has been pointed out by V. Icke (private communication) that in such a
chaotic solution, the switch between modes is made easier if the radial
order $n$ changes by an even number.  He argues that a change between
adjacent modes, such as the fundamental and first overtone, requires the
two modes to be in anti-phase in order to match the velocities at the
stellar surface.  Internally, these two modes will have opposite velocities
over much of the star and the switch would have a large energy requirement.
This transition can therefore be considered as 'forbidden.'  Two modes with
$\Delta n = 2$ are much more similar inside the star and a switch between
them will be energetically favoured.  This simple consideration agrees well
with our suggestion that switching in R~Dor is between the first and third
overtones.

Interestingly, in the \citeone{IFH92} model, R~Dor-like behaviour comes
{\em after\/} stable Mira behaviour, as the envelope mass decreases.  This
agrees with the extended IRAS shell around R~Dor indicating a long
mass-loss history, and also with the behaviour of V~Boo.  It is the only
model in which semiregulars are Mira descendants rather than Mira
progenitors.

\section{Conclusions}

We have shown that R~Dor switches between two pulsation modes, on a time
scale of a few cycles.  The longer mode is shown to fit the Mira P-L
relation extremely well, which shows that semiregulars and Miras are
closely related.  The Hipparcos parallax for R~Dor places it only 62\,pc
away, making it the nearest Mira-like star.  It clearly makes sense to
include long-period semiregulars such as R~Dor in studies of Mira
behaviour.

{}From stellar models, we make a possible identification for the two modes
as the first and third overtones, implying that Mira variables on the P-L
relation pulsate in the first overtone.  The physical diameter of R~Dor
excludes fundamental mode pulsation for reasonable masses.  We suggest that
the change between semiregulars and Miras is due to a mode switch between
third and first overtone, or between higher overtones.  However, by ruling
out fundamental mode pulsation we are left with the problem of large
observed shock velocities.  Until this is explained, the controversy over
the mode of Mira pulsation cannot be laid to rest.

Models giving mode switching through non-linear growth of the fundamental
mode and an adjustment of the thermal structure of the star predict a much
longer time scale for switching than observed.  In other words, the change
in R~Dor occurs too rapidly to be related to a change in the overall
thermal structure of the star.  Chaotic effects discussed by
\citeone{IFH92} appear to fit R~Dor very well.  They predict that irregular
behaviour becomes stronger when the envelope mass of the star decreases.
This agrees with the evidence for extensive past mass loss and implies that
R~Dor, and perhaps other semiregulars, may have evolved from true Miras.

\section*{Acknowledgements}

We appreciate the efforts of Ranald McIntosh and Frank Bateson in
maintaining the RASNZ database.  We thank Peter Wood for the pulsation
calculations reported in Table~\ref{tab.RDor.Wood}.  We also thank him and
Gordon Robertson for comments on the manuscript.  Data for V~Boo were
obtained from the AFOEV database, operated at CDS, France and the VSOLJ
database in Japan.  TRB is grateful to the Australian Research Council for
financial support.  AAZ thanks ESO for financial support.

\if\preprint0
	\clearpage
	\renewcommand{\baselinestretch}{1}
	\begin{table}
\caption[]{\label{tab.RDor.Data} Fundamental parameters of R~Dor}

\begin{tabular}{ll}
\hline
\noalign{\smallskip}
Distance	&  $62.4\pm2.8$\,pc	\\
Radius   	& $383\pm38$\,R$_\odot$	\\
Effective temperature 	& $2740\pm190$\,K	\\
Luminosity 	& $6830\pm900$\,L$_\odot$	\\
$M_{\rm bol}$ 	& $-4.87 \pm 0.13 $		\\
Periods 	& 332 and $\sim$175 days \\
\noalign{\smallskip}
\hline
\end{tabular}
\end{table}

	\renewcommand{\baselinestretch}{2}
	\clearpage
	\begin{figure*}

\centerline{ \includegraphics[draft=false,bb=86 341 541 703]
{figures/rdor-jones-light.ps}}

\caption[]{\label{fig.jones-light} Light curve of R~Dor (observations by
A. Jones).  The dashed curve shows a sinusoid with a period of 332 days.}

\end{figure*}

	\if\preprint0
	\twocolumn
\fi
\begin{figure}

\centerline{ 
\includegraphics[draft=false, bb=63 345 278 719, width=\the\hsize]
{figures/rdor-jones-amp.ps}}

\caption[]{\label{fig.jones-amp} Amplitude spectra of the data in
Fig.~\ref{fig.jones-light}.  The data were taken as a whole (top panel),
and also in two parts (middle and lower panels), where the division was
made at JD\,2434500.   }

\end{figure}

\if\preprint0
	\onecolumn
\fi

	\begin{figure*}

\centerline{
\includegraphics[draft=false, bb=36 122 525 512]
{figures/rasnz-Jo.005.ps}}

\caption[]{\label{fig.wwz.jones005} Wavelet analysis of observations of
R~Dor by A. Jones.  The top panel shows the light curve reproduced from
Fig.~\ref{fig.jones-light}.  The middle panel shows the WWZ and the bottom
panel shows the WWA evaluated at periods of 332\,d (thick line) and 175\,d
(thin line).  See text for more details. }

\end{figure*}

	\begin{figure*}

\centerline{
\includegraphics[draft=false, bb=36 122 525 512]
{figures/sim.005.ps}}

\caption[]{\label{fig.sim} Analysis of simulated data.  The light curve
(top) consists of a long-period component that decays linearly with time
and a short-period component with constant amplitude.  The WWZ (middle)
shows both periods, but give the false impression that the short-period
component is gradually increasing in strength.  Evaluating the WWA (bottom)
gives the correct amplitude behaviour at both the long period (thick line)
and the short period (thin line).}

\end{figure*}

	\begin{figure*}

\centerline{ 
\includegraphics[draft=false,bb=80 125 556 638]
{figures/rasnz-light.ps}}

\caption[]{\label{fig.all-light} Light curve of R Dor from RASNZ data
(including observations in Fig.~\ref{fig.jones-light}).  In the bottom
panel, the diamonds are Tycho $VT$ data, shifted upwards by 1.0\,mag for
readability.  In the top panel, asterisks are maxima and crosses are minima
(in photographic magnitudes, shifted upwards by 0.5\,mag), reported by
\citeone{Gap52}.  Finally, note that all observations after JD\,2450400
were made by A.~Jones.  The dotted sine curve has the same parameters as
that in Fig.~\ref{fig.jones-light}.}

\end{figure*}

	\begin{figure*}

\centerline{
\includegraphics[draft=false,bb=36 122 525 512]
{figures/rdor-all.005.ps}}

\caption[]{\label{fig.wwz.all005} Wavelet transform for R~Dor.  Same as
Fig.~\ref{fig.wwz.jones005}, but based on the data in
Fig~\ref{fig.all-light}.}

\end{figure*}

	\begin{figure}

\centerline{ 
\includegraphics[width=7cm,bb=32 37 288 741,clip]{figures/rdor-mirapl.ps}}

\caption[]{\label{fig.mirapl} The position of R~Dor (filled circle)
relative to the P-L relation for LMC Miras.  }

\end{figure}

	\begin{figure*}

\centerline{
\includegraphics[draft=false,bb=36 122 525 512]
{figures/vboo-both5.005.ps}}

\caption[]{\label{fig.vboo} Wavelet transform for V~Boo based on the AFOEV
and VSOLJ data.  Same layout as Fig.~\ref{fig.wwz.jones005}, except that
the periods in the lower panel (WWA) are 258\,d (thick line) and 137\,d
(thin line).}

\end{figure*}

\else
	\bsp
\fi

\end{document}